\documentclass[3p,times]{elsarticle}
\pdfoutput=1
\usepackage[T1]{fontenc}
\usepackage{graphicx}
\usepackage{float}
\usepackage{epsfig,graphics,subfigure,psfrag,amsmath,amssymb}
\usepackage{lineno}
\usepackage{dcolumn}
\usepackage{bm}
\usepackage{overpic}
\usepackage{xspace}
\usepackage{rotating}
\usepackage{epstopdf}
\usepackage{amsmath}
\usepackage{multirow}
\usepackage{multicol}
\usepackage{bookmark}
\usepackage{hyperref}

\begin{document}

\makeatletter
\begin{frontmatter}

\title{{\bf \boldmath Measurement of the inclusive branching fraction
    for $\psi(3686)\rightarrow K_{S}^{0} + \text{anything}$}}

\author{
M.~Ablikim$^{1}$, M.~N.~Achasov$^{10,b}$, P.~Adlarson$^{64}$, S. ~Ahmed$^{15}$, M.~Albrecht$^{4}$, R.~Aliberti$^{28}$, A.~Amoroso$^{63A,63C}$, Q.~An$^{60,47}$, X.~H.~Bai$^{54}$, Y.~Bai$^{46}$
% ...
\\
\vspace{0.2cm}
(BESIII Collaboration)\\
\vspace{0.2cm} {\it
$^{1}$ Institute of High Energy Physics, Beijing 100049, People's Republic of China\\
% ...
$^{4}$ Bochum Ruhr-University, D-44780 Bochum, Germany\\
% ...
$^{10}$ G.I. Budker Institute of Nuclear Physics SB RAS (BINP), Novosibirsk 630090, Russia\\
% ...
$^{15}$ Helmholtz Institute Mainz, Staudinger Weg 18, D-55099 Mainz, Germany\\
% ...
$^{28}$ Johannes Gutenberg University of Mainz, Johann-Joachim-Becher-Weg 45, D-55099 Mainz, Germany\\
% ...
$^{46}$ Southeast University, Nanjing 211100, People's Republic of China\\
$^{47}$ State Key Laboratory of Particle Detection and Electronics, Beijing 100049, Hefei 230026, People's Republic of China\\
% ...
$^{54}$ University of Jinan, Jinan 250022, People's Republic of China\\
% ...
$^{60}$ University of Science and Technology of China, Hefei 230026, People's Republic of China\\
% ...
$^{63}$ University of Turin and INFN, (A)University of Turin, I-10125, Turin, Italy; (B)University of Eastern Piedmont, I-15121, Alessandria, Italy; (C)INFN, I-10125, Turin, Italy\\
$^{64}$ Uppsala University, Box 516, SE-75120 Uppsala, Sweden\\
% ...
$^{b}$ Also at the Novosibirsk State University, Novosibirsk, 630090, Russia\\
% ...
}
\vspace{0.4cm}
}

\begin{abstract}
Using 5.9~pb$^{-1}$ of $e^+e^-$ annihilation data collected at center-of-mass
energies from 3.640 to 3.701 GeV with the BESIII detector at the
BEPCII Collider, we measure the observed cross sections of
$e^+e^-\rightarrow K_{S}^{0} X$ (where $X= \text{anything}$). From a
fit to these observed cross sections with the sum of continuum and
$\psi(3686)$ and $J/\psi$ Breit-Wigner functions and considering
initial state radiation and the BEPCII beam energy spread, we obtain
for the first time the product of $\psi(3686)$ leptonic width
and inclusive decay branching fraction
$\Gamma^{ee}_{\psi(3686)}\mathcal B(\psi(3686)\rightarrow K_{S}^{0} X)
= (373.8 \pm 6.7 \pm 20.0)$ eV, and assuming
$\Gamma^{ee}_{\psi(3686)}$ is $(2.33 \pm 0.04)$ keV from PDG value, we measure
$\mathcal B(\psi(3686)\rightarrow K_{S}^{0} X) = (16.04 \pm 0.29 \pm 0.90)\%$,
where the first uncertainty is statistical and the second is
systematic.  \\ \\ \text{Keywords: \texorpdfstring{$\psi(3686)$}{��(3686)}, inclusive branching
  fraction, \texorpdfstring{$K_{S}^{0}$}{K\_{S}\textasciicircum{0}}, BESIII}
\end{abstract}

\end{frontmatter}
\makeatother

\begin{multicols}{2}

\section{Introduction}

The decays of $\psi(3686)$ provide an ideal laboratory to study the
strong interaction between the charm quark and antiquark in the low
energy region.  The decay rates of $\psi(3686)$ to some exclusive
final states can be predicted~\cite{psip_theory} by effective theories
based on Quantum Chromodynamics (QCD).  Although the $\psi(3686)$
decays have been studied for more than 40 years since its discovery
in 1974~\cite{psi2S}, the sum of the branching fractions (BFs) for all
the decay channels in the PDG~\cite{PDG2018} is only approximately
90\%, indicating that there are still many decay modes missing. Searching
for new exclusive decay channels and measuring their BFs is important
to test the QCD calculations of $\psi(3686)$ decays, which can lead to
better understanding of the strong interaction in the low energy
region.

Measurements of the BFs of inclusive $\psi(3686)$ decays, which
include transitions, radiative decays, and hadronic decays, can guide
the search for new exclusive decay modes, which could supply missing
BFs for $\psi(3686)$ and other states, such as $J/\psi$ and
$\chi_{cJ}$ ($J=0,1,2$) produced in the $\psi(3686)$
transitions~\cite{psip_JpsiX,psip_gchic}.

The $K_{S}^{0}$ is a long-lived particle, which is easily
reconstructed in the detector, and it can be used as a probe to study
the inclusive decays of $\psi(3686)$.  In this paper, the BF of
$\psi(3686)\rightarrow K_{S}^{0}X$ ($X=\text{anything}$) is measured for the
first time by fitting the observed inclusive $K_{S}^{0}$ cross
sections in $e^{+}e^{-}$ annihilation in the $\psi(3686)$ energy
region.  The line-shape of the $\psi(3686)$ cross section is
described by a Breit-Wigner function, in which the BF is a
parameter~\cite{BWfunc}.

\section{Expected Observed Cross Section}\label{sec:II}

For $e^{+}e^{-}\rightarrow f$, where $f$ are hadronic final states,
the expected observed cross section, taking into consideration the
initial state radiation (ISR) and the beam energy spread to describe
the beam energy resolution, at a center-of-mass (c.m.)  energy
$\sqrt{s}$ is
\begin{linenomath*}
\begin{equation}
\sigma^{\rm exp}(s) = \int_{0}^{\infty}ds'G(s,s')\int_{0}^{1}dx \cdot \sigma^{\rm Dress}(s(1-x))F(x,s),
\label{equ:Xsec2}
\end{equation}
\end{linenomath*}
\noindent where $x$ is the ratio of the total energy of the emitted
photons to the beam energy~\cite{ISR} and $\sigma^{\rm Dress}(s)$ is
the total dressed cross section for $e^{+}e^{-}\rightarrow f$,
which includes the effects of vacuum polarization.  For
the resonances, such as $J/\psi$ and $\psi(3686)$, the dressed cross
section of $J/\psi\rightarrow f$ and $\psi(3686)\rightarrow f$ can be
described by the Breit-Wigner formula; here, exactly the same
parametrisation as given in Ref.~\cite{BWfunc} is used:

\begin{linenomath*}
\begin{equation}
\sigma^{\rm Dress}(s) = \frac{12\pi\Gamma^{ee}\Gamma^{\rm tot}\mathcal B(R \rightarrow f)}{(s - M^{2})^{2} + (\Gamma^{\rm tot}M)^{2}},
\label{equ:Xsec_Rs}
\end{equation}
\end{linenomath*}

\noindent where $M$ and $\Gamma^{\rm tot}$ are the mass and the total
width of the resonance, $\Gamma^{ee}$ is the partial width to the
$e^{+}e^{-}$ channel, and $\mathcal B(R \rightarrow f)$ is the BF for
the resonance decay to the final state $f$.  For the continuum, we assume
that the dressed cross section has the energy dependence defined as

\begin{linenomath*}
\begin{equation}
\sigma^{\rm Dress}(s) = \frac{f_{\rm con}}{s},
\label{equ:Xsec_qqbar}
\end{equation}
\end{linenomath*}

\noindent where $f_{\rm con}$ can be determined experimentally.

$F(x,s)$ is a sampling function based on the structure function
approach by Kuraev and Fadin~\cite{ISR}, given by

\begin{linenomath*}
\begin{equation}
F(x,s) = \beta x^{\beta-1}\delta^{V+S}+\delta^{H},
\label{equ:ISR_function}
\end{equation}
\end{linenomath*}

\noindent where $\beta$ is the electron equivalent radiator thickness,

\begin{linenomath*}
\begin{equation}
\beta = \frac{2\alpha}{\pi}(L-1),
\label{equ:beta}
\end{equation}
\end{linenomath*}

\noindent and

\begin{linenomath*}
\begin{equation}
L = \ln\frac{s}{m_{e}^{2}}.
\label{equ:Len}
\end{equation}
\end{linenomath*}

\noindent Here, $m_{e}$ is the mass of the electron and $\alpha$ is the
fine structure constant.  The correction term for the virtual process and
a soft photon is

\begin{linenomath*}
\begin{equation}
\delta^{V+S} = 1+\frac{3}{4}\beta+\frac{\alpha}{\pi}(\frac{\pi^{2}}{3}-\frac{1}{2})
                +\frac{\beta^{2}}{24}(\frac{37}{4}-\frac{L}{3}-2\pi^{2}),
\label{equ:delta_VandS}
\end{equation}
\end{linenomath*}

\noindent and the correction term for the hard photon is
\begin{equation}
\begin{aligned}
\delta^{H} &= -\beta(1-\frac{x}{2})+\frac{1}{8}\beta^{2}[4(2-x)\ln{\frac{1}{x}}\\
           &  -\frac{1+3(1-x)^{2}}{x}\ln{(1-x)}-6+x].
\label{equ:delta_H}
\end{aligned}
\end{equation}

$G(s,s')$ is a Gaussian function to describe the beam energy
spread. It is defined as

\begin{linenomath*}
\begin{equation}
G(s,s') = \frac{1}{\sqrt{2\pi}\Delta}e^{-(\sqrt{s}-\sqrt{s'})^{2}/(2\Delta^{2})},
\label{equ:Gauss}
\end{equation}
\end{linenomath*}

\noindent where $\Delta$ is the standard deviation of the c.m. energy
distribution, $\sqrt{s}$ and $\sqrt{s'}$ are the nominal and actual
c.m. energies, respectively.

By fitting the observed cross section of $e^{+}e^{-}\rightarrow
K_{S}^{0}X$ as a function of c.m. energy with the sum of the expected
continuum function and $J/\psi$ and $\psi(3686)$ Breit-Wigner
functions, the product of $\Gamma^{ee}_{\psi(3686)}$ and
BF of $\psi(3686)\rightarrow K_{S}^{0}X$ can be
measured and taking PDG~\cite{PDG2018} value for $\Gamma^{ee}_{\psi(3686)}$,
the BF of $\psi(3686)\rightarrow K_{S}^{0}X$ is obtained.
This approach has the advantage compared to directly measuring
the BF at the $\psi(3686)$ resonance, where BESIII has a large data
sample, of unfolding the continuum contribution from the resonance
contribution and allowing the BF and cross section of
$e^{+}e^{-}\rightarrow K_{S}^{0}X$ to be determined at the same time.

\section{Detector and Monte Carlo Simulation}\label{sec:Detector_And_MC}

The BESIII detector is a magnetic spectrometer~\cite{BES3} located at
the Beijing Electron Positron Collider (BEPCII)~\cite{BEPCII}. The
cylindrical core of the BESIII detector consists of a helium-based
multilayer drift chamber (MDC), a plastic scintillator time-of-flight
system (TOF), and a CsI(Tl) electromagnetic calorimeter (EMC), which
are all enclosed in a superconducting solenoidal magnet providing a
1.0 T magnetic field. The solenoid is supported by an octagonal
flux-return yoke with resistive plate chamber muon identifier modules
interleaved with steel. The acceptance of charged particles and
photons is 93\% over the 4$\pi$ solid angle. The charged particle
momentum resolution at 1 GeV/$c$ is 0.5\%, and the specific energy
loss ($dE/dx$) resolution is
6\% for the electrons from Bhabha scattering. The EMC measures photon
energies with a resolution of 2.5\% (5\%) at 1 GeV in the barrel (end
cap) region. The time resolution of the TOF barrel part is 68 ps,
while that of the end cap part is 110 ps.

Three Monte Carlo (MC) simulated data samples (`inclusive MC') of $\psi(3686)\rightarrow
\text{anything}$, $J/\psi\rightarrow \text{anything}$ and $e^{+}e^{-}\rightarrow
q\bar{q}$ $(q = u, d, s)$ have been generated with {\sc
kkmc}~\cite{kkmc} which simulates $\psi(3686)$, $J/\psi$ and
$q\bar{q}$ production in $e^{+}e^{-}$ annihilation, while the
subsequent decays of $\psi(3686)$ and $J/\psi$ are handled by {\sc
evtgen}~\cite{evtgen}.  For the known decay modes, the BFs are set to
the world average values~\cite{PDG2018}, while the remaining unknown
decay modes are modeled by {\sc lundcharm}~\cite{LundCharm} in the
{\sc evtgen} generator. The simulated samples are produced with a
{\sc geant4}-based~\cite{geant4} MC software that includes the geometric
description~\cite{geant4_v1,geant4_v2} of the BESIII detector and the
detector response, and they are reconstructed to determine the
detection efficiency and estimate the backgrounds.  Each sample is
generated with 600000 events for six energy points in the range from
3.645 GeV to 3.697 GeV to determine the efficiency dependence on the
c.m. energy.  The signal MC samples for $\psi(3686)\rightarrow
K_{S}^{0}X$, $J/\psi\rightarrow K_{S}^{0}X$ and $e^{+}e^{-}\rightarrow
K_{S}^{0}X$ are selected with generator information from the inclusive
MC samples of $\psi(3686)$, $J/\psi$ and $e^{+}e^{-}\rightarrow
q\bar{q}$, respectively.  The numbers of events generated for each
signal MC sample are summarized in Table \ref{Tab:MC_samples} for the
six energy points.  To study the backgrounds, an inclusive MC sample
of $1.06\times10^{8}$ $\psi(3686)$ events is used, referred to as ``the
standard $\psi(3686)$ inclusive MC sample'' in the following. We also
use samples of three QED processes, with $e^{+}e^{-}\rightarrow
e^{+}e^{-}$ and $e^{+}e^{-}\rightarrow \mu^{+}\mu^{-}$ both generated
by {\sc babayaga}~\cite{babayaga}, and $e^{+}e^{-}\rightarrow
\tau^{+}\tau^{-}$ generated by {\sc kkmc}~\cite{kkmc}.

\begin{table*}[htb]
\centering
\caption{Numbers of signal MC events.}
  \small
\begin{tabular}{c|c|cccccc}
\hline
\hline
\multicolumn{2}{c|}{$\sqrt{s}$~(GeV)}                                                     & 3.6451 & 3.6534  & 3.6789  & 3.6840  & 3.6860  & 3.6964  \\
\hline
 Signal   &  $\psi(3686)\rightarrow K_{S}^{0}X$                & 111693 & 111337  & 111385  & 111441  & 111326  & 111224  \\
                      MC               &  $J/\psi\rightarrow K_{S}^{0}X$                    & 101611 & 102237  & 101869  & 101489  & 101992  & 101988  \\
                        samples                &  $e^{+}e^{-}\rightarrow K_{S}^{0}X$                & 104595 & 105020  & 105097  & 104830  & 104778  & 105192  \\
\hline
\hline
\end{tabular}
\label{Tab:MC_samples}
\end{table*}

In this analysis we use the $\psi(3686)$ cross-section scan data
collected by BESIII in June 2010 at 22 energy points between 3.640 and
3.701 GeV with a total integrated luminosity of about 5.9 $\rm
pb^{-1}$.  The c.m.\ energies and the corresponding integrated
luminosities are listed in Table \ref{Tab:xs_table}.  In addition, the
data sample of ($106.41\pm0.86)\times 10^{6}$
events~\cite{big_psip_events} collected in 2009 at $\sqrt{s} =
3.686\rm\ GeV$ is also used for a number of studies and is referred to
as ``the standard $\psi(3686)$ data'' in the following.

\begin{table*}[htb]
  \centering
  \vspace{0.05in}
  \small
  \caption{
  The values of the integrated luminosity, $\mathcal L$, the number of observed inclusive $K_{S}^{0}$ events, $N^{\rm obs}$,
  the number of background events, $N^{\rm bkg}$, the detection efficiency, $\epsilon^{e^+e^-\rightarrow K_{S}^{0} X}$ ,
  the observed cross section, $\sigma^{\rm obs}$,
  and the corresponding dress cross section, $\sigma^{\rm dress}$, obtained at each c.m. energy point, $\sqrt{s}$.
  }
  \begin{tabular}{@{\extracolsep{\fill}}crclrcrccrclrcl}
  \hline
  \hline
         \multicolumn{1}{c}{$\sqrt{s}$ (GeV) }
       & \multicolumn{3}{c}{$\mathcal L$ (nb$^{-1}$)}
       & \multicolumn{3}{c}{$N^{\rm obs}$}
       & \multicolumn{1}{c}{$N^{\rm bkg}$}
       & \multicolumn{1}{c}{$\epsilon^{e^+e^-\rightarrow K_{S}^{0} X}$ (\%)}
       & \multicolumn{3}{c}{$\sigma^{\rm obs}$ (nb)}
       & \multicolumn{3}{c}{$\sigma^{\rm dress}$ (nb)}\\
  \hline
    3.6451 &  568.7 & $\pm$ & 2.4 &  345.8 & $\pm$ & 27.1 & 0.0 $\pm$ 0.0 & 23.47 $\pm$ 0.05 &   2.59 & $\pm$ & 0.20 &   2.02 & $\pm$ & 0.16   \\
    3.6474 & 2260.9 & $\pm$ & 4.8 & 1465.3 & $\pm$ & 56.9 & 0.0 $\pm$ 0.0 & 23.47 $\pm$ 0.05 &   2.76 & $\pm$ & 0.11 &   2.15 & $\pm$ & 0.09   \\
    3.6534 & 2217.7 & $\pm$ & 4.8 & 1475.9 & $\pm$ & 55.7 & 0.0 $\pm$ 0.0 & 23.47 $\pm$ 0.05 &   2.84 & $\pm$ & 0.11 &   2.22 & $\pm$ & 0.09   \\
    3.6789 &   49.1 & $\pm$ & 0.7 &   34.6 & $\pm$ & 10.7 & 0.0 $\pm$ 0.0 & 23.31 $\pm$ 0.04 &   3.02 & $\pm$ & 0.93 &   2.60 & $\pm$ & 0.80   \\
    3.6799 &   46.5 & $\pm$ & 0.7 &   13.4 & $\pm$ &  9.6 & 0.0 $\pm$ 0.0 & 23.25 $\pm$ 0.04 &   1.24 & $\pm$ & 0.89 &   1.09 & $\pm$ & 0.78   \\
    3.6809 &   49.6 & $\pm$ & 0.7 &   58.9 & $\pm$ & 10.2 & 0.0 $\pm$ 0.0 & 23.12 $\pm$ 0.04 &   5.14 & $\pm$ & 0.89 &   4.53 & $\pm$ & 0.79   \\
    3.6818 &   52.2 & $\pm$ & 0.7 &   47.0 & $\pm$ &  9.5 & 0.0 $\pm$ 0.0 & 22.87 $\pm$ 0.03 &   3.94 & $\pm$ & 0.80 &   3.07 & $\pm$ & 0.63   \\
    3.6822 &   51.0 & $\pm$ & 0.7 &   70.3 & $\pm$ & 10.9 & 0.1 $\pm$ 0.0 & 22.69 $\pm$ 0.03 &   6.07 & $\pm$ & 0.95 &   4.08 & $\pm$ & 0.67   \\
    3.6826 &   51.2 & $\pm$ & 0.7 &  113.2 & $\pm$ & 13.2 & 0.1 $\pm$ 0.0 & 22.51 $\pm$ 0.04 &   9.82 & $\pm$ & 1.15 &   5.29 & $\pm$ & 0.71   \\
    3.6834 &   51.8 & $\pm$ & 0.7 &  195.9 & $\pm$ & 17.5 & 0.3 $\pm$ 0.0 & 22.25 $\pm$ 0.04 &  16.98 & $\pm$ & 1.54 &   5.47 & $\pm$ & 0.58   \\
    3.6840 &   50.7 & $\pm$ & 0.7 &  418.6 & $\pm$ & 24.3 & 0.7 $\pm$ 0.0 & 22.15 $\pm$ 0.05 &  37.21 & $\pm$ & 2.23 &   9.35 & $\pm$ & 0.63   \\
    3.6846 &   48.7 & $\pm$ & 0.7 &  609.8 & $\pm$ & 29.6 & 1.2 $\pm$ 0.0 & 22.11 $\pm$ 0.05 &  56.50 & $\pm$ & 2.87 &  15.14 & $\pm$ & 0.82   \\
    3.6848 &   39.9 & $\pm$ & 0.6 &  717.9 & $\pm$ & 32.0 & 1.1 $\pm$ 0.0 & 22.10 $\pm$ 0.05 &  81.33 & $\pm$ & 3.86 &  24.74 & $\pm$ & 1.35   \\
    3.6854 &   38.0 & $\pm$ & 0.6 &  875.3 & $\pm$ & 35.2 & 1.5 $\pm$ 0.1 & 22.09 $\pm$ 0.05 & 104.00 & $\pm$ & 4.53 &  79.50 & $\pm$ & 7.26   \\
    3.6860 &   41.2 & $\pm$ & 0.6 &  961.3 & $\pm$ & 36.8 & 1.9 $\pm$ 0.1 & 22.08 $\pm$ 0.05 & 105.52 & $\pm$ & 4.38 &1275.71 & $\pm$ & 160.43 \\
    3.6866 &   40.1 & $\pm$ & 0.6 &  916.0 & $\pm$ & 35.6 & 1.8 $\pm$ 0.1 & 22.09 $\pm$ 0.05 & 103.11 & $\pm$ & 4.34 &  85.80 & $\pm$ & 8.63   \\
    3.6873 &   40.7 & $\pm$ & 0.6 &  748.9 & $\pm$ & 32.5 & 1.4 $\pm$ 0.1 & 22.10 $\pm$ 0.05 &  83.18 & $\pm$ & 3.86 &  20.33 & $\pm$ & 1.13   \\
    3.6874 &   40.1 & $\pm$ & 0.6 &  645.2 & $\pm$ & 30.1 & 1.4 $\pm$ 0.1 & 22.10 $\pm$ 0.05 &  72.65 & $\pm$ & 3.59 &  16.34 & $\pm$ & 0.91   \\
    3.6890 &   40.7 & $\pm$ & 0.7 &  291.0 & $\pm$ & 20.5 & 0.4 $\pm$ 0.0 & 22.20 $\pm$ 0.05 &  32.19 & $\pm$ & 2.33 &   6.65 & $\pm$ & 0.52   \\
    3.6920 &   41.6 & $\pm$ & 0.7 &  107.2 & $\pm$ & 12.9 & 0.1 $\pm$ 0.0 & 22.48 $\pm$ 0.04 &  11.46 & $\pm$ & 1.39 &   3.65 & $\pm$ & 0.45   \\
    3.6964 &   49.7 & $\pm$ & 0.7 &   57.6 & $\pm$ &  9.9 & 0.1 $\pm$ 0.0 & 22.68 $\pm$ 0.03 &   5.11 & $\pm$ & 0.88 &   1.93 & $\pm$ & 0.33   \\
    3.7002 &   50.7 & $\pm$ & 0.7 &   72.7 & $\pm$ & 10.5 & 0.1 $\pm$ 0.0 & 22.80 $\pm$ 0.03 &   6.28 & $\pm$ & 0.91 &   2.64 & $\pm$ & 0.39   \\
  \hline
  \hline
  \end{tabular}
  \label{Tab:xs_table}
\end{table*}

\section{Data Analysis}

\subsection{Measurement of \texorpdfstring{$\sigma^{\rm obs}(e^{+}e^{-}\rightarrow K_{S}^{0}X)$}
                           { ��\textasciicircum{\rm obs}(e\textasciicircum{+}e\textasciicircum{-}�� K\_{S}\textasciicircum{0}X)}}

The $e^{+}e^{-}\rightarrow K_{S}^{0}X$ candidate events, called
inclusive $K_{S}^{0}$ events, are reconstructed using the most
abundant $K_{S}^{0}$ decay to $\pi^{+}\pi^{-}$. More than two good
charged tracks are required with $|\cos\theta| < 0.93$, $R_{xy} <
10\rm\ cm$, and $R_{z} < 20\rm\ cm$, where $\theta$ is the polar angle
with respect to the $z$ axis, while $R_{xy}$ and $R_{z}$ are the
distances of the closest approach to the interaction point in the
plane perpendicular to and along $z$, respectively.  To select the
$K_{S}^{0}$ daughter candidates, good charged tracks are assumed to be
pions, and particle identification is not used. The candidates must
satisfy the following selection criteria: (1) the total charge of the
two tracks is zero; (2) the ratio $E/p$ of each pion candidate is
be less than 0.9 to reject electrons, where $E$ is the energy
deposited in the EMC and $p$ is the momentum reconstructed in the MDC;
(3) for each candidate pair, a secondary vertex fit~\cite{vertex_fit}
is performed; the decay length $L_{\rm decay}$ between the nominal
interaction point and the secondary vertex is required to be larger
than zero, and the combination with longest decay length ($L_{\rm
  max}$) is retained for further analysis. A further requirement,
chosen by optimizing the ratio $S/\sqrt{S+B}$, where $S$ and $B$ are
the numbers of signal and background events estimated from the
standard $\psi(3686)$ inclusive MC sample, $L_{\rm max}>0.4$~cm is
applied.  After the selection of the $K_{S}^{0}$ daughter candidates,
at least one of the remaining good charged tracks is required to
satisfy $R_{xy} < 1\rm\ cm$ and $R_{z} < 10\rm\ cm$.

\begin{figure}[H]
  \centering
  \includegraphics[width=0.4\textwidth]{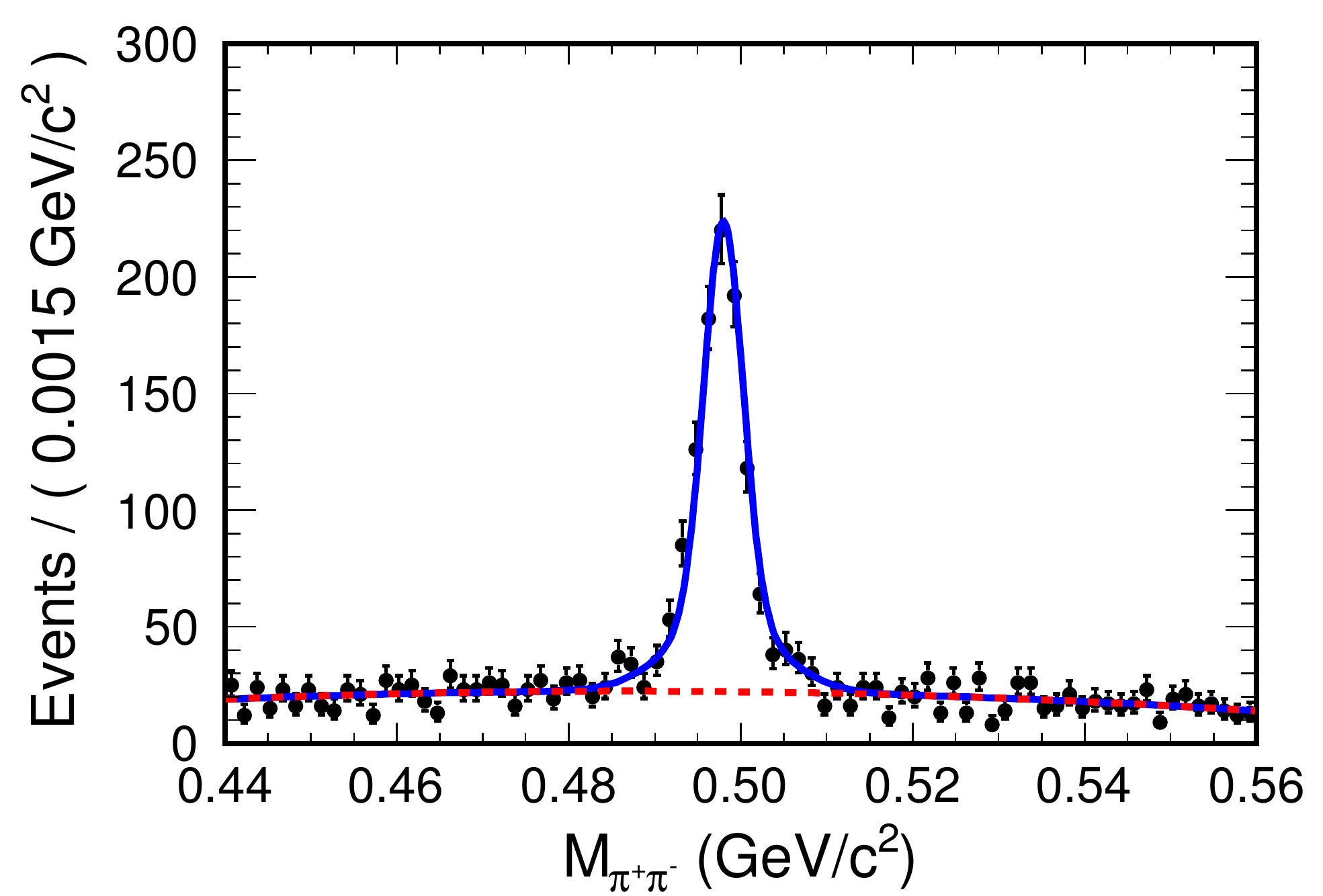}\\
  \caption{
  The $\pi^{+}\pi^{-}$ invariant mass spectrum.
  Points with errors are scan sample data collected
  at $\sqrt{s}$ = 3.686 GeV.
  The blue solid line is the fit result,
  and the red dashed line represents the background contribution.
  }
  \label{fig:Mpipi}
\end{figure}

To obtain the signal yield at each energy point, we perform a maximum
likelihood fit to the invariant mass spectrum of $\pi^{+}\pi^{-}$ with
a Double-Gaussian function and a second-order Chebychev polynomial
function, which are used to describe the signal and background,
respectively. In the fit, the two Gaussian functions have a common
mean value, and their parameters are fixed to the ones obtained by
a fit to the $M_{\pi^{+}\pi^{-}}$ distribution from all data samples
combined. As an example, Fig.~\ref{fig:Mpipi} shows the fit result for
the data set collected at $\sqrt{s} = 3.686$ GeV, where the
$K_{S}^{0}$ signal is clearly seen. The yields of inclusive
$K_{S}^{0}$ events, $N^{\rm obs}$, are obtained for each energy point
and listed in Table \ref{Tab:xs_table}.

The possible sources of background are $e^+e^-\rightarrow
(\gamma)e^+e^-$, $e^+e^-\rightarrow (\gamma)\mu^+\mu^-$,
$e^+e^-\rightarrow \tau^+\tau^-$ and non-$K^0_S$ events from
$\psi(3686)$ decays. Among them, the only peaking background is
$\psi(3686)\rightarrow K_{L}^{0}+\rm Y$, where Y can be anything
except $K_{S}^{0}$. The number of these events ($N^{\rm bkg}$) is
estimated by

\begin{linenomath*}
\begin{equation}
N^{\rm bkg} = \mathcal L \times \eta^{\psi(3686)\rightarrow K_{L}^{0}+\rm Y} \times \sigma^{\psi(3686)\rightarrow K_{L}^{0}+\rm Y},
\label{equ:Sub_Bkg}
\end{equation}
\end{linenomath*}

\noindent where $\mathcal L$ is the integrated luminosity of the data
set and $\eta^{\psi(3686)\rightarrow K_{L}^{0}+\rm Y}$ is the
probability to misidentify the $\psi(3686)\rightarrow K_{L}^{0}+\rm Y$
event as an inclusive $K_{S}^{0}$ event. The value
$\eta^{\psi(3686)\rightarrow K_{L}^{0}+\rm Y} = (6.3 \pm 0.2)\times
10^{-4}$ is estimated from the inclusive MC samples.  The observed
cross section $\sigma^{\psi(3686)\rightarrow K_{L}^{0}+\rm Y}$ is
calculated from Eq.~(\ref{equ:Xsec2}-\ref{equ:Gauss}), with the BF for
$\psi(3686)\rightarrow K_{L}^{0}+\rm Y$ estimated using the standard
$\psi(3686)$ inclusive MC sample, where known decays with $K_L^0$ are
combined with those generated by {\sc lundcharm}~\cite{LundCharm}. At
all energy points the estimated cross section
$\sigma^{\psi(3686)\rightarrow K_{L}^{0}+\rm Y}$ is similar to the
measured $\sigma^{\psi(3686)\rightarrow K_{S}^{0}+\rm X}$.  The
estimated number of peaking background events for each energy is
reported in the fourth column of Table \ref{Tab:xs_table}.

\begin{figure}[H]
  \centering
  \includegraphics[width=0.4\textwidth]{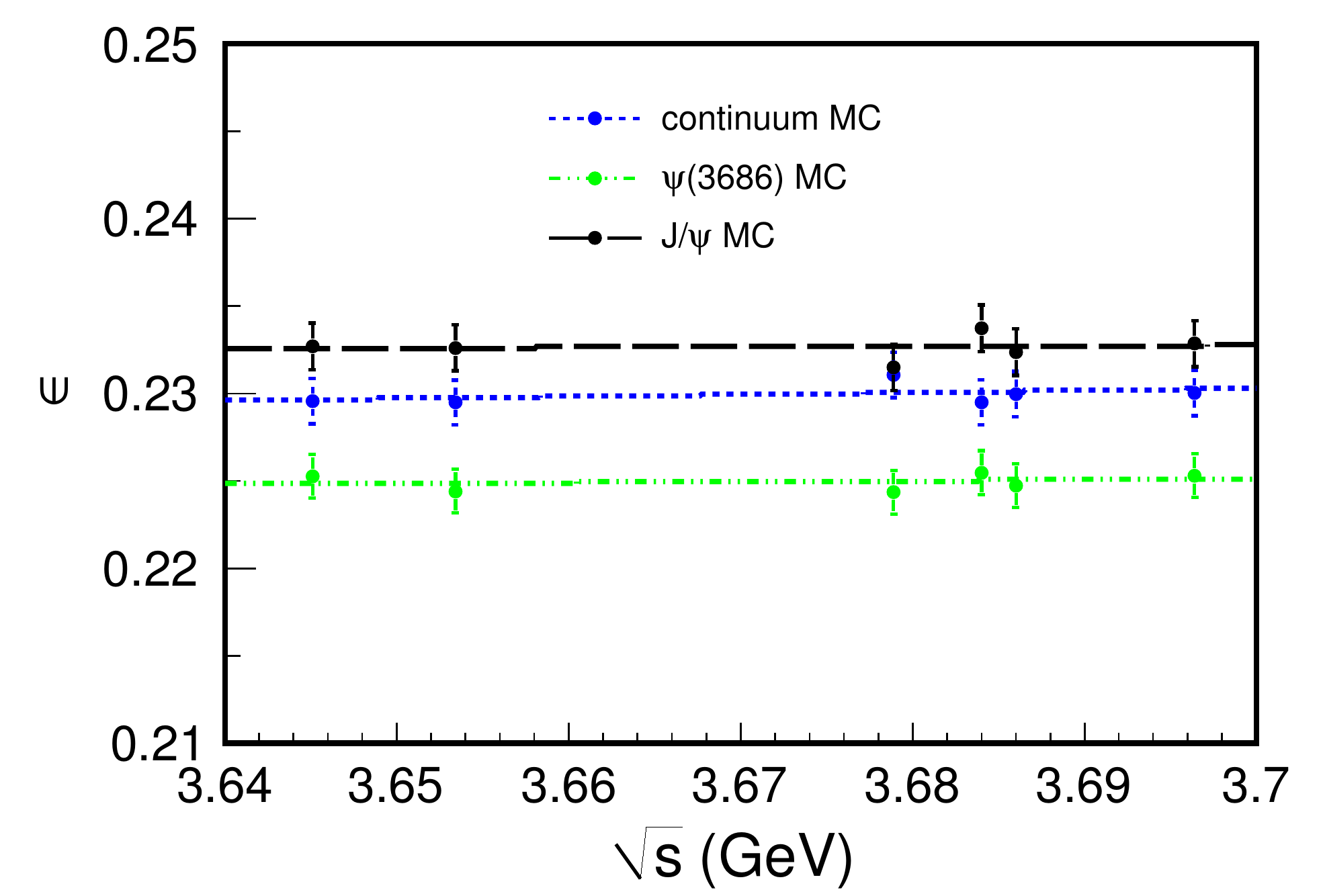}
  \caption{
  Detection efficiency as a function of c.m.\ energy for each
  signal process.
  }
  \label{fig:eff_each_process}
\end{figure}

The detection efficiencies for the three signal processes are
the ratios of the reconstructed events and the total number of
events in the corresponding signal MC samples.  For each signal
process, MC samples at six different c.m.\ energies are generated, and
a linear dependence of the efficiency on $\sqrt{s}$ is found, as
shown in Fig.~\ref{fig:eff_each_process}.  The efficiency values at
other c.m.\ energies are determined by extrapolation of the respective
linear fitting function.
The three signal processes studied differ slightly in the angular
distribution of the inclusive $K_{S}^{0}$.  The efficiency differences visible
in Fig.~\ref{fig:eff_each_process} are caused by the interplay of these angular distributions and
the implicit fiducial cuts applied during event reconstruction.

\begin{figure*}[htb]
  \centering
  \includegraphics[width=0.4\textwidth]{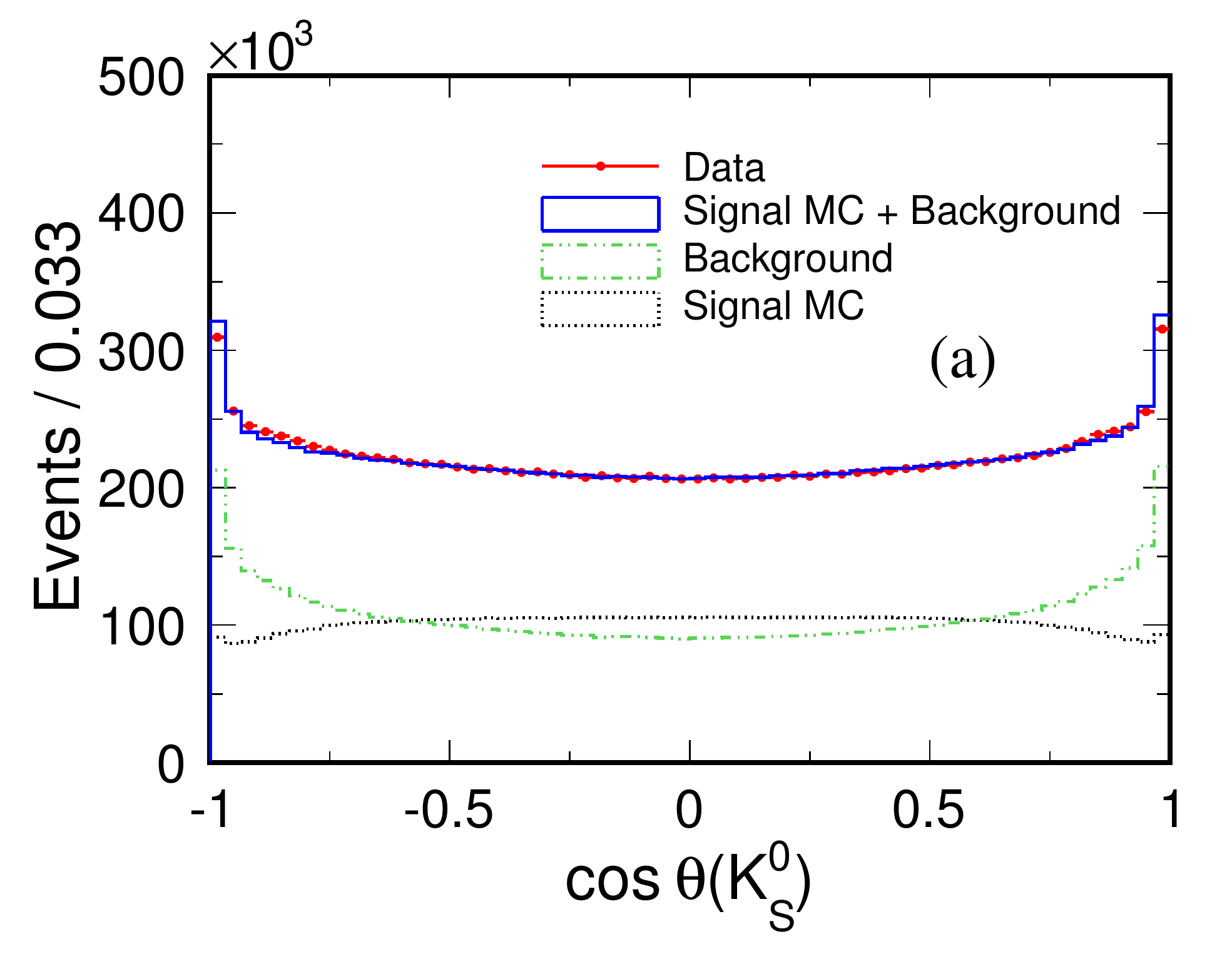}
  \includegraphics[width=0.4\textwidth]{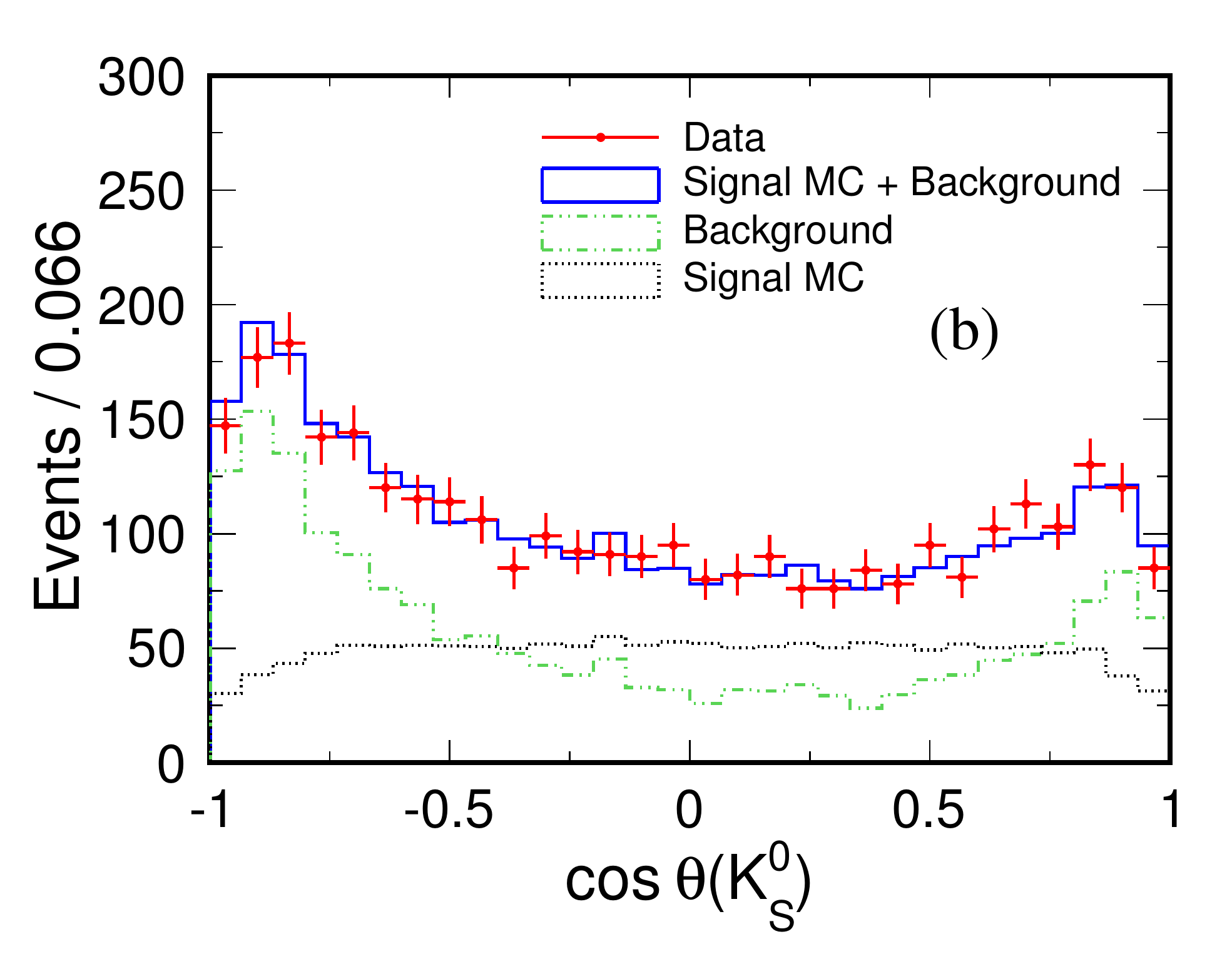}
  \includegraphics[width=0.4\textwidth]{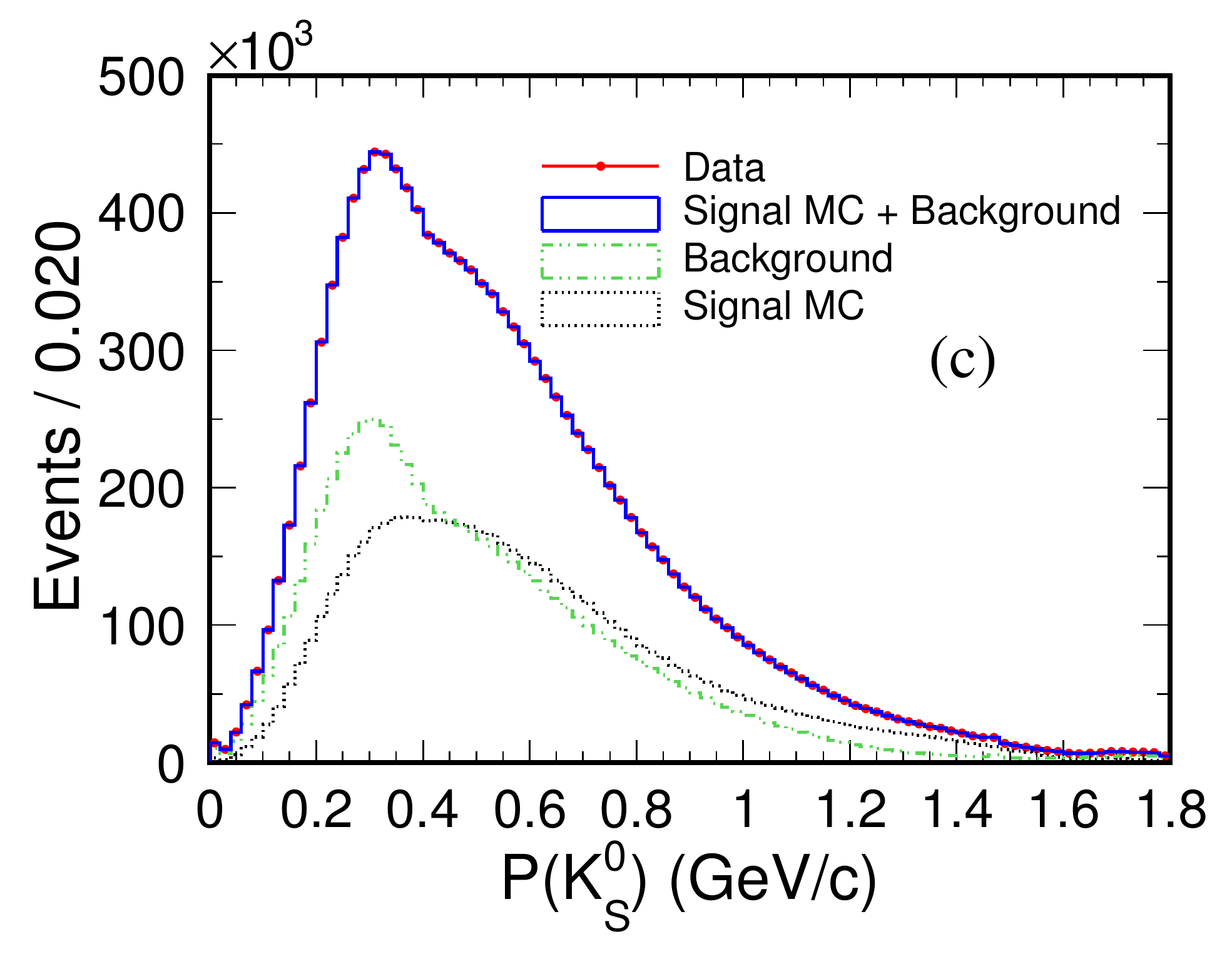}
  \includegraphics[width=0.4\textwidth]{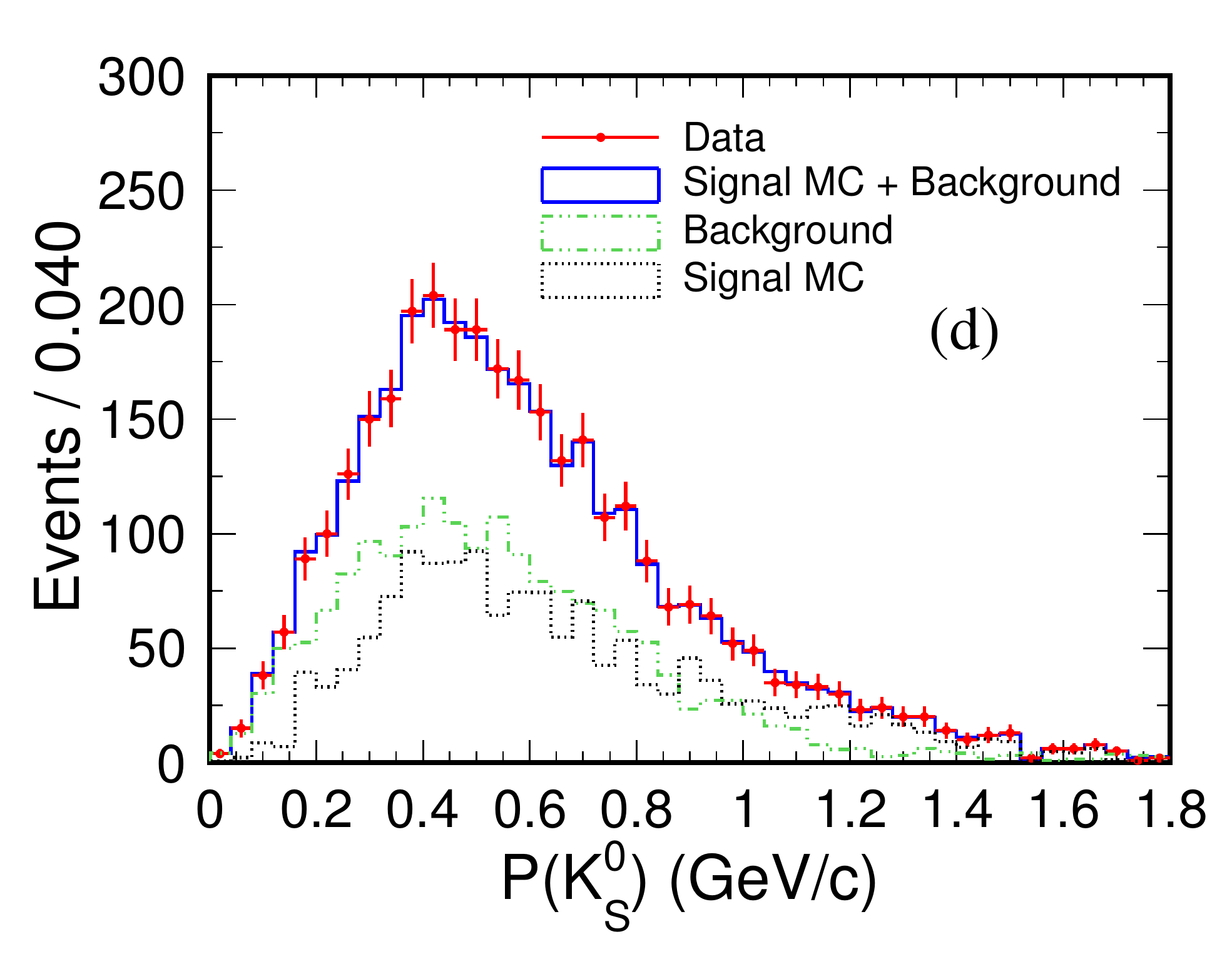}
  \caption{
  Comparisons between the standard $\psi(3686)$ data and the
  standard $\psi(3686)$ MC sample
  at $\sqrt{s} = 3.686\rm \ GeV$ for $K_{S}^{0}$ (a) angular and (c) momentum distributions after corrections.
  The comparisons between continuum data and continuum $K_{S}^{0}X$ MC
  sample at $\sqrt{s} = 3.6534\rm \ GeV$
  for $K_{S}^{0}$ (b) angular and (d) momentum distributions after
  corrections. The red dots with error bars
  are data, the black dotted histograms are the signal MC events, the green dash-ditted histograms are the estimated background
  events and the blue solid histograms are the sum of signal MC events
  and estimated background events.
  }
  \label{fig:cmp_ctht_P_KS0_36534_36861}
\end{figure*}

To improve the reliability of the efficiency estimation, a bin-by-bin
correction is applied to the $K_{S}^{0}$ momentum and angular
distributions in the $\psi(3686)$ and continuum MC samples.  No
correction is made for the $J/\psi$ samples, since the contribution of
ISR $J/\psi$ events is small in the studied energy region. In the
plots (a) and (c) of Fig.~\ref{fig:cmp_ctht_P_KS0_36534_36861}, the
distributions after the correction are shown at $\sqrt{s} =
3.686$~GeV, where the backgrounds are estimated with selection
$M_{\pi^{+}\pi^{-}}\in(0.4, 0.6){\rm \ GeV}/c^{2}$ in the MC sample.
In the same figure, the plots (b) and (d) show the comparisons for
continuum between the data and the MC sample at $\sqrt{s} =
3.6534$~GeV, where the backgrounds are estimated from the data
sidebands $21 < |M_{\pi^{+}\pi^{-}} - M_{K_{S}^{0}}| < 42 {\rm \ MeV}/c^{2}$ and the
signal region is defined as $|M_{\pi^{+}\pi^{-}} - M_{K_{S}^{0}}| < 11
{\rm \ MeV}/c^{2}$, where $M_{K_{S}^{0}}$ is the nominal mass of
$K_{S}^{0}$~\cite{PDG2018}.  Good agreement between data and MC
samples is observed.

The detection efficiency for the inclusive process
$e^{+}e^{-}\rightarrow K_{S}^{0}X$ at the $i^{th}$ c.m. energy is
determined by

\begin{linenomath*}
\begin{equation}
\begin{split}
\epsilon_{e^+e^-\rightarrow K_{S}^{0}X}^{i}=&\frac{1}{\sigma_{\psi(3686)}^{i}+\sigma_{J/\psi}^{i}+\sigma_{\rm con}^{i}}\times\\
&(\sigma_{\psi(3686)}^{i}\cdot\epsilon_{\psi(3686)}^{i}+\sigma_{J/\psi}^{i}\cdot\epsilon_{J/\psi}^{i}\\
&+\sigma_{\rm con}^{i}\cdot\epsilon_{\rm con}^{i}),
\label{equ:eff}
\end{split}
\end{equation}
\end{linenomath*}

\noindent where $\epsilon_{\psi(3686)}^{i}$, $\epsilon_{J/\psi}^{i}$
and $\epsilon_{\rm con}^{i}$ are the efficiencies of the signal
processes determined for the $i^{th}$ energy point,
while $\sigma_{\psi(3686)}^{i}$, $\sigma_{J/\psi}^{i}$ and
$\sigma_{\rm con}^{i}$ are the corresponding signal cross sections
obtained with an iterative procedure by fitting the measured
line-shape with Eqs.~(\ref{equ:Xsec2}-\ref{equ:Gauss}).
For the first iteration, the efficiency is estimated setting the parameters to
the following initial values: $\Delta$ is set to 1.30 MeV
measured~\cite{BEMS} at $\sqrt{s} = 3.686$ GeV, $f_{\rm con}$ to the
value estimated from the continuum data, $\mathcal B(R \rightarrow
K_{S}^{0}X$) ($R = J/\psi, \psi(3686)$) to the values estimated from
the $J/\psi$ and $\psi(3686)$ signal MC samples, and the other
parameters are set to PDG~\cite{PDG2018} values.

For each data sample the measured observed cross section of
$e^+e^-\rightarrow K_{S}^{0} X$ is determined by
\begin{linenomath*}
\begin{equation}
\sigma^{\rm obs}=\frac{N^{\rm obs}-N^{\rm bkg}}{\mathcal L\times\epsilon^{e^+e^-\rightarrow K_{S}^{0} X}},
\label{equation:cs}
\end{equation}
\end{linenomath*}
\noindent where $N^{\rm obs}$ is the number of observed inclusive $K_{S}^{0}$ events,
$N^{\rm bkg}$ is the number of background events,
$\mathcal L$ is the integrated luminosity and
$\epsilon^{e^+e^-\rightarrow K_{S}^{0} X}$ is the
detection efficiency determined according to Eq.~(\ref{equ:eff}).

\begin{figure}[H]
  \centering
  \includegraphics[width=0.4\textwidth]{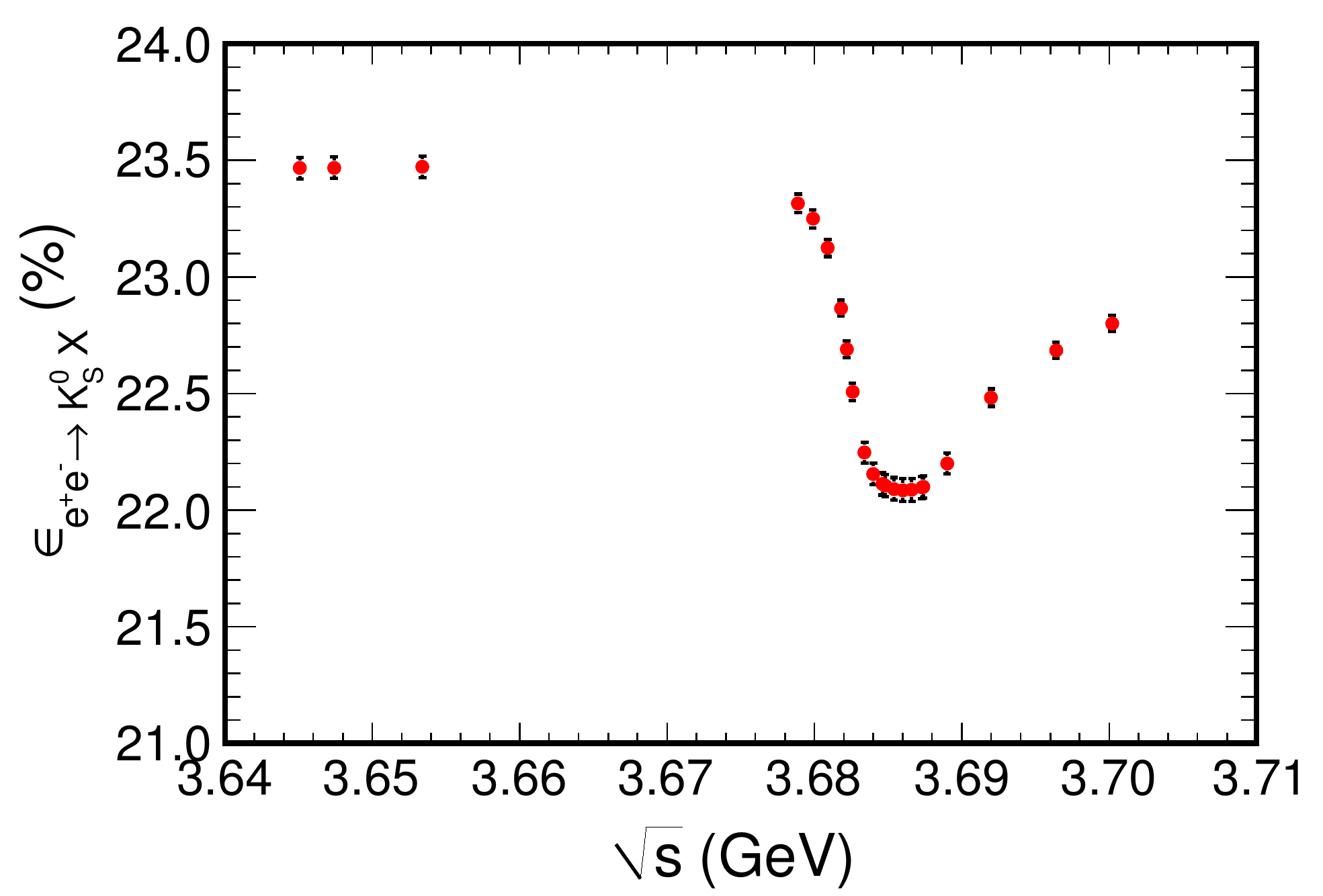}
  \caption{
    Detection efficiency as a function of c.m.\ energy for
    $e^+e^-\rightarrow K_{S}^{0} X$,
    where the vertical axis is expanded.
  }
  \label{fig:Weight_eff_correction}
\end{figure}

The observed cross section at each energy point is first obtained with
the initial detection efficiency.  By fitting the observed cross
sections with Eq.~(\ref{equ:Xsec2}), the parameters of
Eqs.~(\ref{equ:Xsec_Rs}) and (\ref{equ:Xsec_qqbar}) are updated, and new
detection efficiencies are calculated.  The iterations are repeated
until the change of the parameters is less than 0.1\%.
The procedure converges after three iterations.
The expected contribution from $J/\psi$ is around 0.2~nb, while the
continuum contribution varies from 2.54~nb to 2.47~nb across the
energy range. The final detection efficiency for
$e^{+}e^{-}\rightarrow K_{S}^{0}X$ is shown in
Fig.~\ref{fig:Weight_eff_correction} as a function of the c.m.\ energy.
The efficiency values $\epsilon^{J/\psi}_{K_{S}^{0} X}$,
$\epsilon^{\psi(3686)}_{K_{S}^{0} X}$, and $\epsilon^{\rm con}_{K_{S}^{0}
X}$ are estimated to be 23.27\%, 22.05\%, 23.54\%, respectively,
with little variation over the studied energy range.
Since the efficiency contribution of $\psi(3686)$ gradually increases and decreases
on either side of $\psi(3686)$ peak energy and the $\psi(3686)$ efficiency is lower than
the continuum efficiency, the final detection efficiency with energy dependency
in Fig.~\ref{fig:Weight_eff_correction} looks like a valley. The values of the
detection efficiency and the observed cross sections of
$e^{+}e^{-}\rightarrow K_{S}^{0}X$ are listed in Table
\ref{Tab:xs_table}, where only statistical errors are given.

\subsection{Fit to \texorpdfstring{$\sigma^{\rm obs}(e^{+}e^{-}\rightarrow K_{S}^{0}X)$}
                                  { ��\textasciicircum{\rm obs}(e\textasciicircum{+}e\textasciicircum{-}�� K\_{S}\textasciicircum{0}X)} }

We perform a chi-square fit to the energy dependent observed
cross section determined with the iterative procedure described in the
previous section. The fit allows the determination of the product of
$\Gamma^{ee}_{\psi(3686)}$ and BF of $\psi(3686)\rightarrow K_{S}^{0} X$
as described in section \ref{sec:II}.  The total expected cross sections for
$e^+e^-\rightarrow K_{S}^{0} X$ can be written as

\begin{linenomath*}
\begin{equation}
\sigma_{K_{S}^{0}X}^{\rm obs}(s) = \sigma_{K_{S}^{0}X}^{\psi(3686)}(s) + \sigma_{K_{S}^{0}X}^{J/\psi}(s)
                                  + \sigma_{K_{S}^{0}X}^{\rm con}(s),
\label{equ:Xsec_Tot}
\end{equation}
\end{linenomath*}

\noindent which is the sum of the observed cross sections of the three
main processes contributing to the final state.  Any interference
between resonant and continuum contribution is expected to be
different in different specific channels, and is therefore expected to
be negligible in the measurement of the inclusive process.

The parameters of the fit function can be divided into three
groups: $J/\psi$ parameters, $\psi(3686)$ parameters, and the remaining
parameters.  All the $J/\psi$ parameters are fixed to the
PDG~\cite{PDG2018} values except for the $\mathcal B(J/\psi\rightarrow
K_{S}^{0} X)$ which is fixed to the value estimated from the inclusive
MC due to the lack of experimental measurements.  Parameters of
$\psi(3686)$ are free parameters of the fit, except for
$\Gamma^{\rm tot}_{\psi(3686)}$ which
is fixed to the PDG values. Other parameters, $f_{\rm con}$ of
Eq.~(\ref{equ:Xsec_qqbar}) and the beam energy spread $\Delta$ are
free parameters.  The value of $\Delta$ is assumed to be constant in
the whole energy range used in the fit.

\begin{figure}[H]
  \centering
  \includegraphics[width=0.4\textwidth]{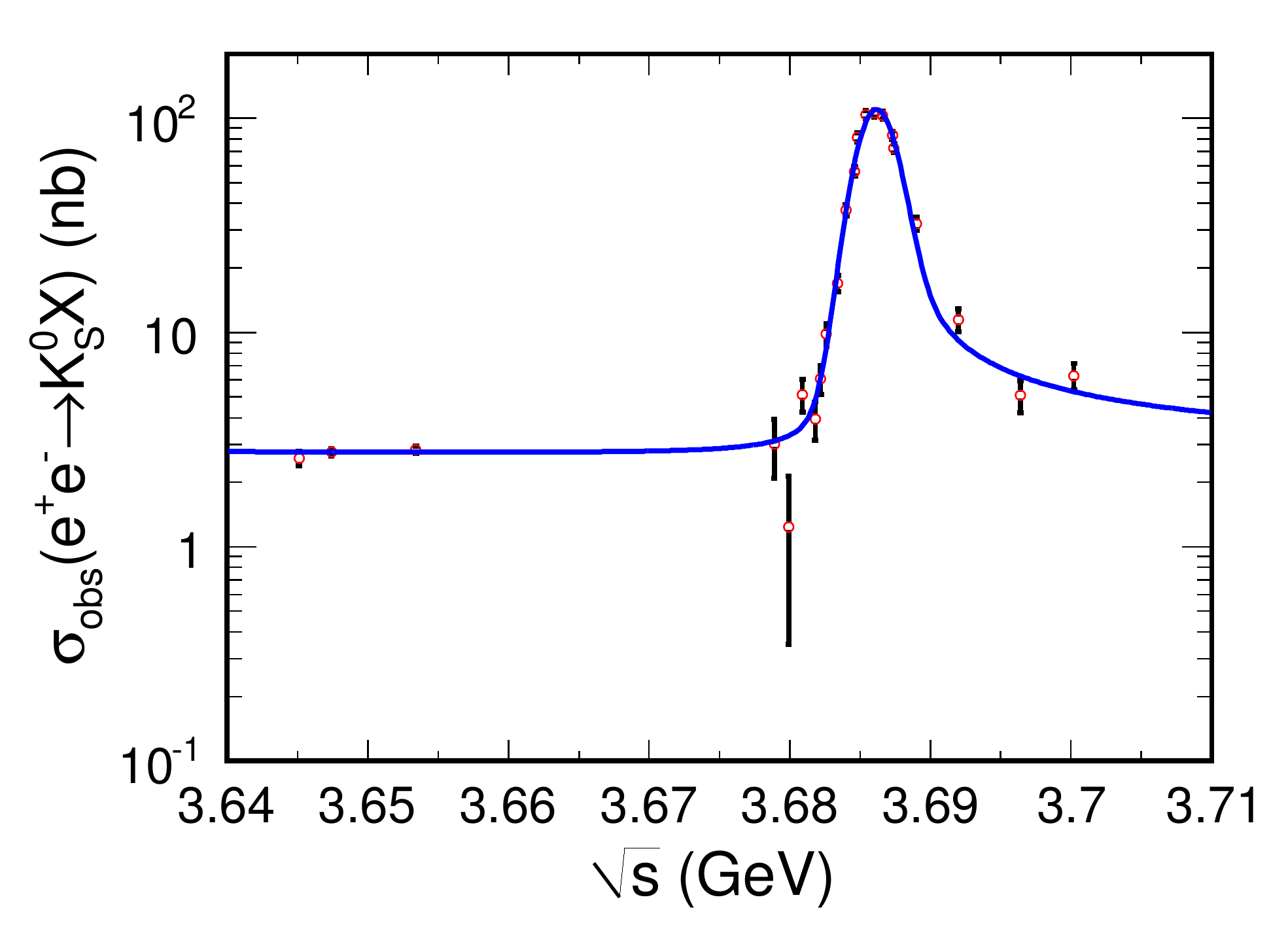}
  \caption{
  Best fit to the observed cross sections for $e^{+}e^{-} \rightarrow K_{S}^{0} X$.
  The red dots with error bars are the measured cross sections, and
  the blue line is
  the fit result.
  }
  \label{fig:fitcs}
\end{figure}

The fit is performed using only the statistical uncertainties of the
measured cross sections.  The best fit result is shown in
Fig.~\ref{fig:fitcs}.  The product of $\Gamma^{ee}_{\psi(3686)}$
and BF for the inclusive decay of $\psi(3686)\rightarrow K_{S}^{0}X$
is determined to be
$\Gamma^{ee}_{\psi(3686)}\mathcal B(\psi(3686)\rightarrow K_{S}^{0}X)
= (373.8 \pm 6.7)$ eV. In Table\ref{table:fitcs}, the parameters
of the best fit function are summarized. The large $\chi^{2}/n.d.f$ is
dominated by the two points at the c.m.\ energies of 3.6848 and 3.6854 GeV.
Without these points the $\chi^{2}/n.d.f$ is $27.2/16$ and
$\Gamma^{ee}_{\psi(3686)}\mathcal B(\psi(3686)\rightarrow K_{S}^{0}X)$
changes to 365.8 eV. Assuming $\Gamma^{ee}_{\psi(3686)} = (2.33 \pm 0.04)$ keV
~\cite{PDG2018}, the BF for the inclusive decay of
$\psi(3686)\rightarrow K_{S}^{0}X$ is measured to be
$\mathcal B(\psi(3686)\rightarrow K_{S}^{0}X) = (16.04 \pm 0.29)\%$.

\begin{table}[H]
  \centering
  \small
  \caption{Results of the fit to the observed cross sections for $e^{+}e^{-} \rightarrow K_{S}^{0} X$.}
  \begin{tabular}{lc}
  \hline
  \hline
  Parameter                                                                       &  Solution \\
  \hline
  Energy spread [MeV]                                                             &  $1.33 \pm 0.03$    \\
  $f_{\rm con}$                                                                   &  $28.49 \pm 0.80$   \\
  $M_{J/\psi}$ [MeV$/c^2$]                                                        &  3096.9  (fixed)    \\
  $\Gamma^{\rm tot}_{J/\psi}\rm[keV]$                                             &  92.9    (fixed)    \\
  $\Gamma^{ee}_{J/\psi}\mathcal B(J/\psi\rightarrow K_{S}^{0}X)\ \rm[eV]$         &  941.28  (fixed)    \\
  $M_{\psi(3686)}$ [MeV$/c^2$]                                                    &  $3686.03 \pm 0.03$ \\
  $\Gamma^{\rm tot}_{\psi(3686)}\rm[keV]$                                         &  294  (fixed)       \\
  $\Gamma^{ee}_{\psi(3686)}\mathcal B(\psi(3686)\rightarrow K_{S}^{0}X)\ \rm[eV]$ &  $373.8 \pm 6.7$    \\
  $\chi^{2}/n.d.f$                                                                &  45.3/18             \\
  \hline
  \hline
  \end{tabular}
  \label{table:fitcs}
\end{table}

\section{Systematic Uncertainties}

The systematic uncertainties on
$\Gamma^{ee}_{\psi(3686)}\mathcal B(\psi(3686)\rightarrow K_{S}^{0} X)$
originate mainly from the measurement of the observed cross sections
for $e^+e^-\rightarrow K_{S}^{0} X$, the fitting procedure and the
uncertainties of the c.m. energy.

The systematic uncertainties for the measurement of the observed cross
sections arise from the following sources: (1) event selection
($N_{\rm good}$, $R_{xy}$ and $R_{z}$ requirements); (2) $K_{S}^{0}$
reconstruction; (3) uncertainty of the integrated
luminosity, 1.00\% from Ref.~\cite{lljiang_lum}; (4) uncertainty on
$\mathcal B(K_{S}^{0}\rightarrow \pi^{+}\pi^{-})$, 0.07\% from the
PDG~\cite{PDG2018}; (5) fit to $M_{\pi^{+}\pi^{-}}$; (6) MC modeling;
and (7) background subtraction.  For the sources (1), (5), (6) and
(7), the uncertainties are evaluated by re-measuring the cross section
with the changes described below. The largest deviations from the nominal
results are taken as systematic uncertainties.

For source (1), the selection requirements are changed from
$N_{\rm good} > 2$ to $N_{\rm good} \geq 2$, and for the tracks not
originating from $K_{S}^{0}$ from $R_{xy} < 1.0 \rm\ cm$ and
$R_{z} < 10.0 \rm\ cm$ to $R_{xy} < 10.0 \rm\ cm$ and $R_{z} < 20.0 \rm\ cm$.
The measured observed cross section changes by 0.67\% and 2.87\%,
respectively.  Source (2) is studied using a control sample of
$J/\psi \rightarrow K^*(892)^\pm K^*(892)^\mp$ and
$K^*(892)^\pm \rightarrow K_{S}^{0}\pi^{\pm}$ events collected at
$\sqrt{s} = 3.097\rm\ GeV$. The estimated uncertainty is 2.09\%.
To study source (5) we vary the fit range, the bin width, the signal shape and the
background shape.  Adding the changes of the result in quadrature, the
uncertainty is 3.32\%.  For source (6), the selection efficiency is
evaluated from the MC samples excluding events where the
$K_{S}^{0}$ did not originate from a $K^{0}$ meson according to the
generator information.  The observed change of cross section, 1.51\%,
is taken as the systematic uncertainty.  For source (7), the
$K_{L}^{0} + \rm Y$ production cross section is replaced by the
measured $K_{S}^{0} + \rm X$ cross section, and the change of the BF
by 0.04\% is taken as the systematic uncertainty.

The relative uncertainties in the measurement of the observed cross
section are listed in Table \ref{tab:systematic_xs} for all
sources described above.  Assuming they are independent, the total
value of 5.23\% is obtained by adding them in quadrature according to
the "offset method"~\cite{offset}.

The systematic uncertainties of the fit procedure arise from the
fixed parameters, the continuum parametrization and the energy spread
$\Delta$.  They are evaluated by repeating the fitting procedure with
changes detailed below and taking the difference with the nominal
result of $\Gamma^{ee}_{\psi(3686)}\mathcal B(\psi(3686)\rightarrow K_{S}^{0}X)$
as the systematic uncertainty. For the fixed parameters the largest
uncertainties is from $\Gamma^{\rm tot}_{\psi(3686)}$.  By changing the
parameter by $\pm 1\sigma$ of the PDG error~\cite{PDG2018} the systematic uncertainty is
evaluated to be 0.12\%.  The contribution from the fixed $J/\psi$
parameters is negligible.  The uncertainty from the continuum
parametrization is evaluated by changing the expression of
Eq.~(\ref{equ:Xsec_qqbar}), to $\sigma_{K_{S}^{0}X}^{\rm con} = f_{\rm con}/s^{n}$
and repeating the fit procedure with $n$ as a free
parameter.  The resulting systematic uncertainty is found to be
0.06\%.  For the beam energy spread $\Delta$ the uncertainty is
estimated replacing the value obtained from the fit by the nominal
value 1.30 MeV obtained with the beam energy measurement
system~\cite{BEMS}. The change of 0.68\% in
$\Gamma^{ee}_{\psi(3686)}\mathcal B(\psi(3686)\rightarrow K_{S}^{0}X)$ is
assigned as the systematic uncertainty.

The systematic uncertainty due to the c.m.\ energy is estimated by
changing the energy values within the errors~\cite{BEMS} and
re-fitting the observed cross sections.  The uncertainty is estimated
to be 0.87\%.

Assuming that all the contributions listed above are independent, the
total uncertainty for measuring
$\Gamma^{ee}_{\psi(3686)}\mathcal B(\psi(3686)\rightarrow K_{S}^{0}X)$
is estimated by adding them in quadrature and is found to
be 5.35\%, as summarized in Table \ref{tab:systematic_tot}.

\begin{table}[H]
  \centering
  \small
  \caption{Systematic uncertainties for measuring
  the observed cross sections ($\sigma^{\rm obs}$) of
  $e^{+}e^{-} \rightarrow K_{S}^{0} X$ in \%.
  }
  \begin{tabular}{lc}
  \hline
  \hline
  Source                                                                       &  Systematic uncertainty\\
  \hline
  $N_{\rm good}$                                                               &  0.67\\
  $R_{xy} < 1.0 \rm\ cm$ and $R_{z} < 10.0 \rm\ cm$                            &  2.87\\
  $K_{S}^{0}$ reconstruction                                                   &  2.09\\
  $\mathcal L$                                                                 &  1.00\\
  $\mathcal B(K_{S}^{0} \rightarrow \pi^{+}\pi^{-})$                           &  0.07\\
  Fit to $M_{\pi^{+}\pi^{-}}$                                                  &  3.32\\
  MC modeling                                                                  &  1.51\\
  Background subtraction                                                       &  0.04\\
  \hline
  Total                                                                        &  5.23\\
  \hline
  \hline
  \end{tabular}
  \label{tab:systematic_xs}
\end{table}

\begin{table}[H]
  \centering
  \small
  \caption{Systematic uncertainties for measuring the product of $\Gamma^{ee}_{\psi(3686)}$
  and branching fraction of $\psi(3686) \rightarrow K_{S}^{0} X$ in \%.
  }
  \begin{tabular}{lc}
  \hline
  \hline
  Source                                                                       &  Systematic uncertainty\\
  \hline
  $\sigma^{\rm obs}$                                                           &  5.23\\
  Fixed fit parameters                                                         &  0.12\\
  Continuum parametrization                                                    &  0.06\\
  $\Delta$                                                                     &  0.68\\
  $\sqrt{s}$                                                                   &  0.87\\
  \hline
  Total                                                                        &  5.35\\
  \hline
  \hline
  \end{tabular}
  \label{tab:systematic_tot}
\end{table}

\section{Summary}

The observed cross sections for $e^+e^-\rightarrow K_{S}^{0} X$ (where
$X= \text{anything}$) are measured at 22 energy points in the range from
3.640 to 3.701 GeV using the data collected by BESIII detector at the
BEPCII Collider.  By fitting the observed cross sections as a function
of the c.m. energy, the product of $\Gamma^{ee}_{\psi(3686)}$ and
BF of $\psi(3686)\rightarrow K_{S}^{0} X$ is measured for the first time to be
$\Gamma^{ee}_{\psi(3686)}\mathcal B(\psi(3686)\rightarrow K_{S}^{0}X)=(373.8 \pm 6.7 \pm 20.0)$ eV,
and assuming $\Gamma^{ee}_{\psi(3686)} = (2.33 \pm 0.04)$ keV~\cite{PDG2018},
the BF of $\psi(3686)\rightarrow K_{S}^{0} X$ is determined to be

\begin{linenomath*}
\begin{equation}
\mathcal B(\psi(3686)\rightarrow K_{S}^{0}X)=(16.04 \pm 0.29 \pm 0.90)\%,
\end{equation}
\end{linenomath*}

\noindent where the first uncertainty is statistical and the second is
systematic, combining the error of $\Gamma^{ee}_{\psi(3686)}$ into the
systematic error.  The sum of all the BFs of $\psi(3686)$ decays to
exclusive $K_{S}^{0}$ final states including the transitions followed
by $J/\psi$ and $\chi_{cJ}$ ($J=0,1,2$) decays is $\sim 5.95\%$ as
reported in the PDG~\cite{PDG2018}, which is much lower than the current
measurement.  This suggests that there are many undiscovered exclusive
channels for $\psi(3686)$ decay to final states containing $K_{S}^{0}$.

\section{Acknowledgments}

The BESIII collaboration thanks the staff of BEPCII and
the IHEP computing center for their strong support. This
work is supported in part by National Key Basic Research
Program of China under Contract No. 2015CB856700,
2009CB825204; National Natural Science Foundation of
China (NSFC) under Contracts Nos. 11625523, 11635010,
11735014, 11822506, 11835012, 11961141012, 10935007;
the Chinese Academy of Sciences (CAS) Large-Scale Scientific
Facility Program; Joint Large-Scale Scientific Facility Funds
of the NSFC and CAS under Contracts Nos.
U1532257, U1532258, U1732263, U1832207; CAS Key Research Program
of Frontier Sciences under Contracts Nos.
QYZDJ-SSW-SLH003, QYZDJ-SSW-SLH040; 100 Talents
Program of CAS, CAS Other Research Program under Code
No. Y129360; INPAC and Shanghai Key Laboratory for
Particle Physics and Cosmology; ERC under Contract No.
758462; German Research Foundation DFG under Contracts
Nos. Collaborative Research Center CRC 1044, FOR 2359;
Istituto Nazionale di Fisica Nucleare, Italy; Ministry of
Development of Turkey under Contract No. DPT2006K-120470;
National Science and Technology fund; STFC (United Kingdom);
The Knut and Alice Wallenberg Foundation (Sweden)
under Contract No. 2016.0157; The Royal Society, UK under Contracts
Nos. DH140054, DH160214; The Swedish
Research Council; U. S. Department of Energy under Contracts
Nos. DE-FG02-05ER41374, DE-SC-0010118, DE-SC0012069.

\end{multicols}

\end{document}